\documentclass[12pt]{iopart}
\usepackage[x11names]{xcolor}
\usepackage{mathabx, wasysym, graphicx}

\begin{document}
\title[]{Dark matter searches using accelerometer-based networks}



\author{Nataniel L. Figueroa$^{1,2}$, Dmitry Budker$^{1,2,3}$, Ernst M. Rasel$^{4}$}


\address{$^1$ Johannes Gutenberg-Universit{\"a}t Mainz, 55128 Mainz, Germany\\ $^2$ Helmholtz-Institut, GSI Helmholtzzentrum f{\"u}r Schwerionenforschung, 55128 Mainz, Germany \\ $^3$ Department of Physics, University of California, Berkeley, California 94720, USA\\ $^4$ Leibniz Universit{\"a}t Hannover, Institut f{\"u}r Quantenoptik, Welfengarten 1, D-30167
Hannover, Germany}

\ead{figueroa@uni-mainz.de}

\begin{abstract}
Dark matter is one of the biggest open questions in physics today. It is known that it interacts gravitationally with luminous matter, so accelerometer-based searches are inherently interesting. In this article we present recent (and future) searches for dark matter candidates such as feebly interacting matter trapped inside the Earth, scalar-matter domain walls and axion quark nuggets, with accelerometer networks and give an outlook of how new atomic-interferometry-based accelerometer networks could support dark matter searches.
\end{abstract}

%
\vspace{2pc}

\section{Introduction}

After decades of research, the evidence for dark matter (DM) has become increasingly significant, indicating that $\sim85$\,\% of gravitating matter in the Universe is nonluminous. The nature of DM is still unknown and is one of the greatest unsolved mysteries in physics to this day. DM observations to date have been astronomical, but there is abundant DM in our galaxy, so Earth-bound DM-sensitive experiments should be feasible and could provide key insights into DM's nature and morphology.

It is known that DM interacts gravitationally, so gravity-based DM searches are inherently interesting. Additionally, certain well-motivated DM-particle candidates would generate forces on luminous matter through exotic interactions. Both cases would provide signatures suitable for accelerometer-based detection. This prospect is particularly interesting because networks of gravimeters and seismometers are already in place around Earth and have been collecting data for decades. In this article, we review recent searches for DM using these networks and explore how future searches might look like. Additionally, we discuss the development of accelerometers based on atom-interferometry and their potential use as a network to search for DM. Alternative gravity-based DM searches using lunar ranging (see Ref. \cite{Zhang2020} and references therein) are possible, as well as purely gravitational detection of individual DM particles, which was recently discussed in Ref.\,\cite{Carney2020}; and also \cite{Carney2020a} in this Special Issue.


\section{Superconducting gravimeters}
Superconducting gravimeters \cite{Hinderer2015,Prothero1968} are accelerometers that use superconductors to magnetically levitate a mass, effectively creating a stable spring-mass system. The position of the mass can be monitored and kept at a fixed position using feedback coils. This feedback signal is proportional to the displacement of the test mass, providing a practical method for measuring its acceleration. Such gravimeters have a high precision ($\sim$0.01\,$\mu$Gal; 1\,$\mu$Gal = 10 nm/s$^2$), small drift ($\sim\mu$Gal/year), and have been of great service to the study of a plethora of geophysical phenomena, as well as fundamental research, for example in the search for Lorentz violation in gravity \cite{Shao2018}.

The International Geodynamics and Earth Tide Service (IGETS) \cite{Voigt2016} provides long-term records of superconducting gravimeter stations around Earth aiming to monitor the temporal variations of the gravitational field of the Earth to support geophysical research. The IGETS is a continuation of the Global Geodynamics Project \cite{Crossley2009, Crossley2010} that began operations in 1997, and has amassed a large collection of data that are readily available.


Recently, the IGETS data were used to look for various kinds of DM structures \cite{Horowitz2020a, Hu2020, Mcnally2020}. Although these structures were not observed, useful constraints on their parameters were placed. In the following subsections we discuss the DM structures that were looked for, their expected signature and future possibilities of improved searches.

\subsection{Compact DM object search}
\label{sec:CDOSearch}

This subsection is a summary of the work presented in Refs.\,\cite{Horowitz2020a, Hu2020}.

Many DM models predict the formation of compact DM objects (CDOs) \cite{Liebling2012, Giudice2016, Kimball2018, Yang2017}. The CDOs considered here are assumed to have weak nongravitational interactions, and exclude black holes, as these would devour Earth. Limits for CDOs come from gravitational-lensing searches that have ruled out DM being made of CDOs with masses between $10^{-11}$ and $15$\,Sun\,masses,\,$M_{\tiny \astrosun }$ \cite{Niikura2019}.

\begin{figure}
    \centering
    \includegraphics[width = \textwidth]{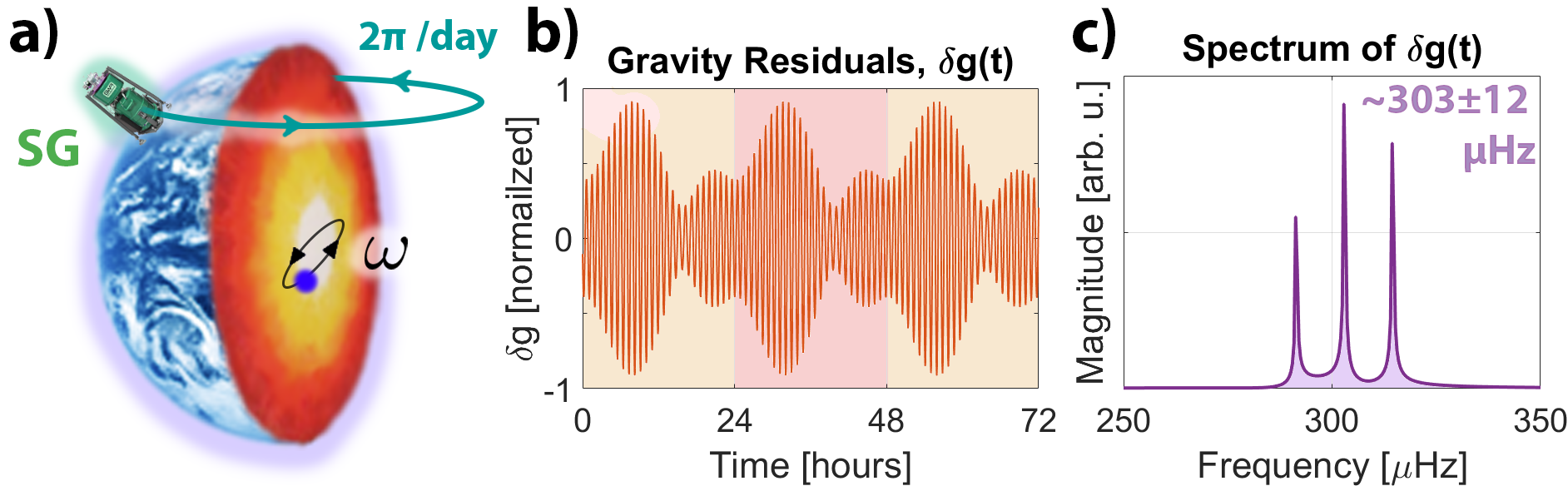}
    \label{fig:CDO}
    \caption{ Superconducting gravimeter search scheme of section \ref{sec:CDOSearch}. The trajectories of a trapped CDO inside of Earth and of a superconducting gravimeter (SG) station on the Earth's surface are shown in (a). The expected gravity residuals and their spectrum as measured by the the station are illustrated in (b) and (c) respectively. Note that the gravity residuals have a fast oscillation (the CDO frequency $\omega / 2\pi \sim$303\,$\mu$Hz), and a slow daily modulation due to the movement of the station. This gives the CDO  spectrum sidebands at $\pm 1$/day\,$\sim$11.6\,$\mu$Hz \cite{Note2}. The linewidths of the spectrum shown in (c) arise from considering a one-month-long window of data.}
    \label{fig:cdo}
\end{figure}

Throughout the lifetime of the Earth, a CDO could have been captured by the planet's gravitational well, through a three-body interaction or some other mechanism \cite{Note1}. Once inside the Earth, the motion of the CDO would gradually dampen by weak nongravitational interactions and/or dynamical friction \cite{Ostriker1999} (which is gravitational in nature), until the CDO's orbit is near the Earth's core (Fig. \ref{fig:cdo}). Then, the orbits will have a well defined period of $2\pi /\omega \approx 55$\,min, as the density of the Earth's core is uniform, and this frequency will be independent of the CDO's mass and the specific geometry of the orbit. A superconducting gravimeter station on the surface of the Earth would experience acceleration due to 1- the direct gravitational attraction of the CDO and 2- the acceleration of the Earth towards the CDO, where these contributions, 1 and 2, are of the same order. The magnitude of the acceleration at the gravimeter station due to the CDO will be approximately
\begin{equation}
    \delta g \approx \bigg( 2+ \frac{\rho_c}{\bar{\rho}} \bigg) \bigg( \frac{m}{M_\Earth} \bigg) \bigg( \frac{\ell}{R_{\Earth}} \bigg)g, \end{equation}
where $\bar{\rho}$ and $\rho_c$ are the average densities of the Earth and the Earth's core, $M_\Earth$ and $m$ are the masses of the Earth and the CDO, $R_\Earth$ and $\ell$ are the radii of the Earth and the CDO's orbital motion, and $g$ is the gravitational acceleration due to the Earth. For a sense of scale, consider that for $m = 10^{-11} M_\Earth$ and $\ell = 0.1 R_\Earth$, $\delta g$ corresponds to $22$\,pm/s$^2$. Note that additional factors are needed to properly account for the dependence of $\delta g$ on the position of the gravimeter with respect to the CDO orbit, a useful discussion  regarding this can be found in the Supplemental Material of Ref.\,\cite{Horowitz2020a}. 

Because of the Earth's rotation, the spectrum of the CDO-induced acceleration at a gravimeter station will acquire sidebands at $\omega \pm (2 \pi /$day$)$. The ratios between the three relevant peak heights will depend on the specifics of the CDO orbit \cite{Note2}. Additionally, inhomogeneites in the density of the Earth lead to higher harmonics in the spectrum, and although their contribution is negligible for oscillations near the center of the Earth, they become relevant for larger CDO amplitudes and could be useful for future searches.

A slightly different approach was taken by Hu and colleagues \cite{Hu2020}. They extracted the residual gravity data from the IGETS network's level 3 data \cite{Note5, Boy2019}, separated it into uninterrupted 1-month-long blocks and averaged their power spectral densities. No excess power was observed at the expected frequency $\omega$, which allowed them to place limits at the $m < 1.3\times10^{-11} M_{\Earth} =8 \times 10^{13} $\,kg level (assuming an $\ell= 0.1 R_\Earth$ orbit). Although more stations were involved in this analysis (providing a more uniform sensitivity across the Earth), the limit is weaker due to the poor scaling of power averaging and the chopping of the data into 1-month-long blocks, which artificially broadens the spectrum of the CDO.

Future Earth-bound searches for CDOs could look into non-conventional CDO orbits, including those not trapped by the Earth but passing through it. Additionally, a more involved analysis of the entire data provided by the IGETS network could already provide an improvement on the current limits. 

\subsection{Domain-wall search}
\label{sec:DWSearch}
This subsection is a summary of the work presented in Ref.\,\cite{Mcnally2020}.

Axions are promising DM candidates, they were originally conceived to solve the strong-CP problem, and it was later found that they could also account for the DM content of the Universe. Axions and other ultralight (masses between $10^{-22}$ and $1$\,eV \cite{Ferreira2020}) DM fields might form spatially inhomogeneous structures that could be arranged in 2D (sheet-like), 1D (line-like) and 0D (point-like) configurations.

As the Earth moves in the galaxy, it would travel through these structures. Furthermore, if we assume that the DM field couples with scalar couplings to luminous matter, travelling through these structures could effectively cause an apparent shifting of fundamental constants. This could lead to a change in the mass of fundamental particles. In particular, for a quadratic coupling, the effective mass of fermions will be given by
\begin{equation}
\label{eq:coupling}
    m_{f}^{*} = m_f(1+ \Gamma_f \phi(r,t)^2),
\end{equation}
where $m_f$ refers to fermion mass, where the fermion could be electron, proton, or neutron; $\Gamma_f$ is the corresponding coupling constant, and $\phi$ is the DM field.

\begin{figure}
    \centering
    \includegraphics[width = \textwidth]{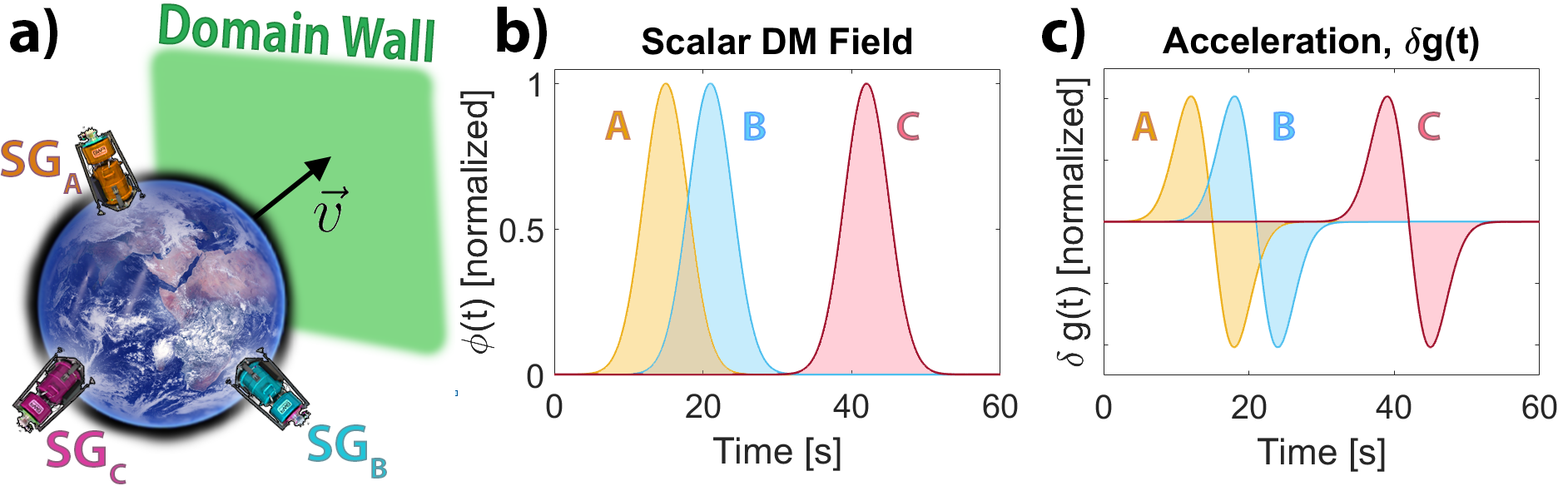}
    \caption{Domain wall acceleration signature, as explained in section \ref{sec:DWSearch}. As Earth travels through a domain-wall (a), the scalar DM field will change at the superconducting gravimeter stations (b). This will generate a transient acceleration (c). Note that the shape of the transient will depend on the width of the wall.}
    \label{fig:domainwall}
\end{figure}

The change of mass of fundamental particles would change the mass of everything in the planet. While a consistent dynamic theory describing the scalar DM interaction with luminous matter is still under development \cite{Note4}, in a naive picture this would lead to a detectable acceleration through the following mechanism: the energy of particles on Earth will be dominated by their rest mass, $E \sim mc^2$, and the coupling described in Eq.\,\ref{eq:coupling} will give mass a spatial dependence. So there will be a force due to the gradient of the energy $\vec{F} = - \vec{\nabla} E$, leading to a corresponding acceleration given by

\begin{equation}
    \vec{a} = \frac{1}{m^{*}(r,t)} \bigg[ - \vec{\nabla}\bigg( {m^{*}(r,t)c^2} \bigg)  \bigg].
\end{equation}
This acceleration transient will affect the test mass in superconducting gravimeters, providing a direct approach for a domain wall search, as illustrated in Fig. \ref{fig:domainwall}.

In a recent article by McNally and Zelevinsky \cite{Mcnally2020}, the authors estimated the sensitivity of the IGETS network to such transient events (for 2D sheet-like structures), and showed that such a search could look for a wide range of structure sizes spanning from $10^{-5}$ to $10^8$\,km. In the case of non-detection, such a search could provide the strongest limits on the coupling constants to date, currently held by a GPS-clock-based search \cite{Roberts2017}.


\section{Atom interferometry based sensors and networks}

Since the famous experiments \cite{Colella1975} exploiting neutrons, one of the most fascinating applications of matter-wave interferometry is probing gravity. The driving force for advancing these interferometers is more stringent and complementary tests of special and general relativistic effects, detecting infrasound gravitational waves and the search for dark matter and dark energy \cite{Tino2020,Elder2016,Chiow2018,Chiow2020}. Moreover, next to fundamental research, applications of matter wave interferometers range from navigation to geophysics and geodesy, for example for monitoring mass-transport phenomena in hydrology or volcanic research. Mass-transport phenomena occur at a wide range of frequencies and include diurnal periods as well as slow processes such as the post-glacial landlift, to mention an extreme case.

Light-pulse atom interferometry \cite{Kasevich1992} is the most common method employed for those measurements including those done with commercial intruments. Light-pulse interferometers belong to the class of absolute gravimeters without need for calibration. They have shown convincing performance with respect to the measurement sensitivity at low frequencies, long-term stability and accuracy. Moreover, light-pulse interferometry offers high flexibility in creating instruments with a wide range of topologies that can measure different quantities like inertial forces, gradients and curvatures, with a single device. 

The most basic type of interferometer, called Mach-Zehnder interferometer in analogy with optics, measures accelerations acting on matter waves during their free fall. The interferometer is created by driving optical, Raman or Bragg processes in the matter wave by stationary or running optical lattices such that the matter wave is coherently split, deflected and recombined by three successive light pulses ideally resulting in two outputs  with complementary interference fringes. The optical lattice is created out of a light wave being reflected off of a mirror serving as inertial reference and connecting the measurement to the laboratory system. The phase depends on the double differential of the atomic motion between the third and the second and the second and the first light pulses. Hence, in the case of a spatially uniform force and constant lattice, the phase $\phi$ is determined by the acceleration $g$, the square of the time $T$ the atoms spend between the light pulses and the relative momentum $k$ of the split matter waves which is proportional to the photon recoil of the corresponding lattice, 

\begin{equation}
    \phi = \vec{k}\cdot \vec{g} \, T^2.
\end{equation}

Starting with the famous Kasevich-Chu interferometer \cite{Kasevich1992}, a large number of such devices were developed around the world since the nineties, especially to measure gravity. These interferometers are presently commercialised by several companies as gravimeters.  Currently, the best measurements reach an accuracy of 32 nm/s$^2$ \cite{Fang2016} and show long-term instabilities of 0.5 nm/s$^2$, necessitating superconducting gravimeters for verification \cite{Freier2016} of the instrument-related instabilities. The use of ultracold atomic ensembles, e.g. those created by evaporative cooling \cite{Karcher2018} or even Bose-Einstein condensates \cite{Heine2020}, promises to improve the results by more than an order of magnitude.

Extending the time of free fall between the light pulses as well as the momentum transferred by the lattice allows to enhance the sensitivity of the interferometer. This motivated current activities building large light-pulse interferometers as pioneered at Stanford \cite{Peters1999}. Moreover, in the context of gravitational wave detection, pairs of such interferometers interacting with the same optical lattice, are under construction forming highly sensitive gravity gradiometers. Most recently, within this device, a quantum test of the equivalence principle, comparing the relative motion of atomic ensembles comprising $^{85}$Rb and $^{87}$Rb has been achieving a sensitivity for differential accelerations of 10$^{-12}$ of the Earth gravity \cite{Asenbaum2020}, improving on previous experiments \cite{Bonnin2013,Zhou2015}. Other isotope-pairs that have also been used to measure differential accelerations are $^{39}$K with $^{87}$Rb \cite{Schlippert2014}, and $^{88}$Sr with $^{87}$Sr \cite{Tarallo2014}. 

Individual atomic interferometers are already sensitive to ultralight DM \cite{Arvanitaki2018,Graham2016,Geraci2016}; but a global network of such sensors could be used to search for DM structures on a larger sale, such as the domain walls described in the previous section. This is an exciting prospect, given that today there are several activities around the world constructing very-long-baseline in Australia\,\,\cite{Hardman2016}, China\,\,\cite{Zhou2011}, and in the US\,\,\cite{Overstreet2018}, as well as gradiometers, where atom interferometers are compared with a common lattice beam covering large distances. These devices will range in size from $\sim$10\,m to $\sim$100\,m with future plans to reach 1000\,m in some cases, and are being constructed in China (ZAIGA)\,\cite{Zhan2020}, France\,(MIGA)\,\cite{Canuel2018}, and the US (MAGIS)\,\cite{Graham2017} or are planned such as the AION\,\cite{Badurina2020} or the European large facility (ELGAR)\,\cite{Bouyer2018}.

\begin{figure}
    \centering
    \includegraphics[width=0.5 \textwidth]{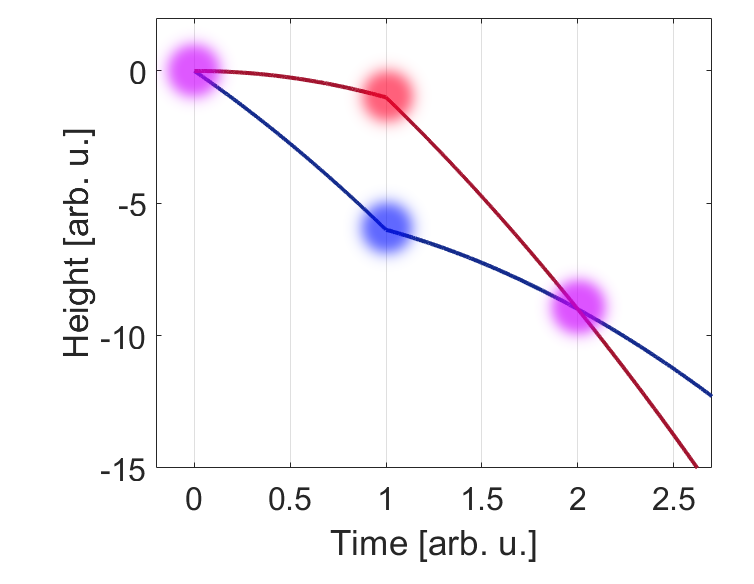}
    \caption{The principle of atomic interferometry. At t=0, an optical pulse prepares an atomic wavepacket (purple cloud) into a superposition of states with different momenta, which separate spatially (red and blue clouds). An optical pulse at t=1 swaps the momentum of the wavepackets. At t=2 the wavepackets are recombined and the phase acquired between both arms of the interferometer is measured.}
    \label{fig:AIFig}
\end{figure}

These atomic interferometers would be sensitive to the class of DM described in the previous section; if there is a different DM coupling to protons and neutrons, it would appear as a differential gravitational acceleration dependent on the proton/neutron ratio of different isotopes. A network of atomic interferometers could be used to look for the domain walls discussed above.

In addition, experiments in space are proposed for testing fundamental physics. Most recently, in the context of ESA’s call for mission ideas for the Voyage 2050 program, several experiments based on atom interferometers were proposed to perform a quantum test of the equivalence principle, for detection of gravitational waves, and to search for dark matter. We note that a comprehensive review of atom interferometry and its utility for DM searches by G. M. Tino can be found in Ref. \cite{Tino2020} in this Special Issue.

\section{Antiquark nugget seismology}
\label{sec:AQNSeismology}

An intriguing possibility is that axion domain walls are not long-lived, and collapse during the QCD transition in the early universe. Domain walls could exhibit substructure at the QCD scale that would prevent fermions from easily crossing them. The collapse of bubble-shaped domain walls would compress the (anti)quarks inside them into a dense state, until an equilibrium is reached when the Fermi pressure of the trapped (anti)quarks counters the squeezing caused by the collapsing domain wall. The pressure exerted by the domain wall could be such that the internal (anti)quarks enter the color-superconducting phase, which would make the system stable, as the mass of these composite objects--known as (anti)quark nuggets--would be smaller than that of their free separated components \cite{Zhitnitsky2003}. 

(Anti)quark nuggets are a compelling DM candidate. Their origin is the same as that of luminous matter (baryons), so they would naturally occur in a comparable abundance without the need of finely-tuned parameters, explaining why we have comparable amounts of dark and luminous matter. Additionally, an excess of antiquark nuggets over quark nuggets can explain the observed matter to antimatter excess in the Universe: matter and antimatter could have been created in the same amount, so the total baryon charge of the Universe would be zero at all times, but more antimatter could be trapped inside nuggets (in contrast to the conventional baryogenesis paradigm where more matter than antimatter was produced in the early Universe). The excess of anti-nuggets to nuggets could arise from strong CP violation in the QCD epoch.

Furthermore, (anti)quark nuggets have been proposed as an explanation for a wide variety of open questions in physics such as: the primordial lithium puzzle \cite{Flambaum2019}, the anomalous temperature of the solar corona \cite{Raza2018}, the creation of small craters that are not attributed to meteorites \cite{Vandevender2020}, some astronomical x-ray emissions \cite{Ge2020}, and skyquakes \cite{Budker2020} which have been routinely observed on Earth.

DM made of (anti)quark nuggets could interact with luminous matter but would remain cosmologically dark due the small cross-section-to-mass ratio of the nuggets. Antiquark nuggets (AQNs) are particularly reactive, and as they travel through luminous matter, part of the AQN contents would annihilate with the medium and emit x-ray photons. If this were to happen inside the Earth, the released energy would generate a seismic signature that could be looked for with seismometers. It should be noted that seismic searches looking for related DM candidates \cite{Witten1984, Madsen1998} have been already carried out in the Earth \cite{Anderson2003} and Moon \cite{Herrin2006}.

The seismic signature of an AQN would have properties that would make it discernible from run-of-the-mill earthquakes.
Assuming that  AQNs are virialized \cite{Green2012} their velocity in the galactic frame along each of the Cartesian coordinates would be distributed as a Gaussian centered at zero, with a dispersion of $\sigma \sim$110\,km/s. As the Earth moves with a velocity of $\sim$220\,km/s through this AQN-gas, the AQN impacts would come from a preferential direction, and with a mean velocity of $\sim$220\,km/s \cite{Lawson2019}.

\begin{figure}
    \centering
    \includegraphics[width = \textwidth]{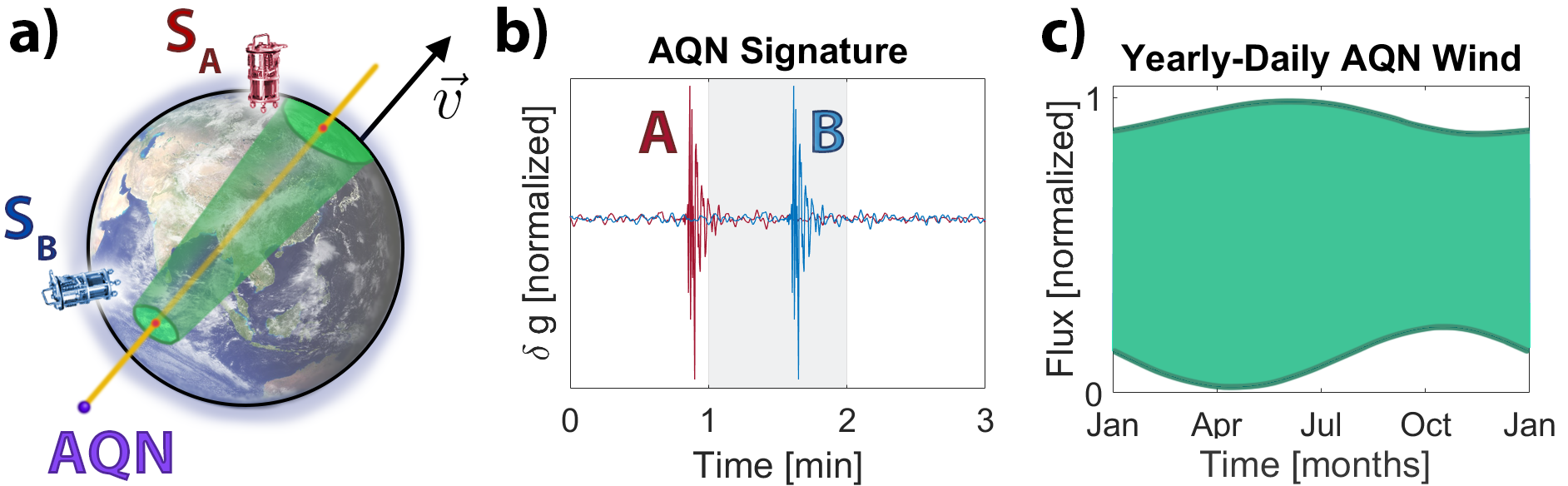}
    \caption{AQN search as explained in section \ref{sec:AQNSeismology}. As an AQN travels supersonically through the Earth (a), it creates seismic waves that are detected by seismometer stations A and B (b), almost simultaneously. The frequency of such events will depend on the DM-wind direction and magnitude, which changes daily and yearly in a predictable fashion as shown in (c), which was calculated for Mainz (50.0$^{\circ}$\,N, 8.2$^{\circ}$\,E) for 2020-2021. Note that (c) is  oscillating daily between the minimum and maximum values shown, but these daily-oscillations are so fast that the plot looks solid.}
    \label{fig:nugget}
\end{figure}

As an AQN travels through the Earth at a supersonic velocity, it generates acoustic waves with a conical, almost cylindrical wavefront (Fig. \ref{fig:nugget}). It also creates surface waves at the points of entry and exit, which would seem to appear nearly simultaneously (within $\sim$1\,min) on opposing sides of the Earth. Furthermore, the preferred direction and frequency of AQN events would have a daily and yearly modulation due to the rotation and motion of the Earth, providing yet another feature that would be exclusive to AQN-produced earthquakes. AQN events would also emit axions \cite{Lawson2019,Budker2020a} that could, in principle, be detected by other networks (but that is a much harder task than detecting sound, as axions would couple weakly to luminous matter).

The sensitivity of a seismometer network to AQN events depends on the search strategy. Designing a promising search strategy is not trivial, especially considering that some noise sources observed by seismometers exhibit daily and yearly modulation of their own \cite{Nakata2019}. Additionally, a proper model for the AQN acoustic generation mechanism must be developed that would consider the complexities of acoustic propagation through the Earth. An ongoing effort is focused on simulating the seismic signature of an AQN event that would inform a search using a seismometer network the near future. Such a search might be able to probe an unexplored range of the free parameter of the AQN model, the average baryon charge in the nuggets, $\langle B \rangle$. Preliminary results from these simulations \cite{Note6} suggest that an AQN with $\langle B \rangle \sim 10^{25}$ could produce a displacement-wave of several nanometers in the vicinity of 10s of meters in solid granite, which is within the detection capabilities of modern seismometers. Larger AQNs would produce bigger displacements, but also occur at a lower rate.

Current constraints on $\langle B \rangle$ come from the IceCube Observatory (see Appendix A in Ref.\,\cite{Lawson2019}), placing a limit of $3\times 10^{24} < \langle B \rangle $ and from a reinterpretation \cite{Budker2020} of the search using the Apollo data mentioned earlier \cite{Herrin2006}, giving $\langle B \rangle < 10^{28}$. Together, they leave an unexplored window $3\times 10^{24} < \langle B \rangle < 10^{28}$, which could be investigated with a seismometer-network.



\section{Conclusion}

DM detection based on the existing and emerging gravimeters and seismometers,  while  being somewhat ``on the fringe'' of DM searches is nevertheless gaining momentum, with good prospects for improved sensitivity and a reach across a broad range of DM models, from ultralight scalar fields to antiquark nuggets. With a broad network of DM detectors pushing technological limits, there are good chances of uncovering the nature and composition of DM in the near future.

\section*{Acknowledgements}

The authors are grateful for discussions with: C. J. Horowitz, R. McNally and T. Zelevinsky regarding their work; W. Hu, M. M. Lawson, D. F. J. Kimball, A. P. Mills Jr. and C. Voigt about the superconducting gravimeter search; F. Tillman, C. Sens-Sch\"onfelder and A. Zhitnitsky with respect to AQN seismology; as well as with H.\,M\"uller, G. M. Tino,  G. Perez and A. Banerjee.

N.L.F. is thankful for the support of the Mainz Physics Academy (part of the PRISMA$^+$ Cluster of Excellence).

D.B. acknowledges support by the Cluster of Excellence Precision Physics, Fundamental Interactions, and Structure of Matter (PRISMA+ EXC 2118/1) funded by the DFG within the German Excellence Strategy (Project ID 39083149).


\section*{References}

\nocite{*}
\bibliographystyle{unsrt}
\bibliography{BIB}

\begin{thebibliography}{10}

\bibitem{Zhang2020}
M.~Zhang, J.~M{\"u}ller, and L.~Biskupek.
\newblock Test of the equivalence principle for galaxy's dark matter by lunar
  laser ranging.
\newblock {\em Celestial Mechanics and Dynamical Astronomy}, 132:25, 2020.

\bibitem{Carney2020}
D.~Carney, S.~Ghosh, G.~Krnjaic, and J.~M. Taylor.
\newblock Proposal for gravitational direct detection of dark matter.
\newblock {\em Physical Review D}, 102(7):072003, 2020.

\bibitem{Carney2020a}
D.~Carney, G.~Krnjaic, D.~C. Moore, C.~A. Regal, G.~Afek, S.~Bhave,
  B.~Brubaker, T.~Corbitt, J.~Cripe, N.~Crisosto, et~al.
\newblock Mechanical quantum sensing in the search for dark matter.
\newblock {\em Quantum Science and Technology}, 2020.

\bibitem{Hinderer2015}
J.~Hinderer, D.~Crossley, and R.~J. Warburton.
\newblock Superconducting gravimetry.
\newblock In {\em Treatise on Geophysics}, pages 59--115. Elsevier, 2015.

\bibitem{Prothero1968}
W.~A. Prothero and J.~M. Goodkind.
\newblock A superconducting gravimeter.
\newblock {\em Review of Scientific Instruments}, 39(9):1257--1262, 1968.

\bibitem{Shao2018}
C.-G. Shao, Y.-F. Chen, R.~Sun, L.-S. Cao, M.-K. Zhou, Z.-K. Hu, C.~Yu, and
  H.~M{\"u}ller.
\newblock Limits on {L}orentz violation in gravity from worldwide
  superconducting gravimeters.
\newblock {\em Physical Review D}, 97(2):024019, 2018.

\bibitem{Voigt2016}
C.~Voigt, C.~F\"{o}rste, H.~Wziontek, D.~Crossley, B.~Meurers, V.~Pálinkáš,
  J.~Hinderer, J.~Boy, J.~Barriot, and H.~Sun.
\newblock Report on the data base of the international geodynamics and earth
  tide service ({IGETS}).
\newblock {\em Scientific Technical Report STR - Data; 16/08}, 2016.

\bibitem{Crossley2009}
D.~Crossley and J.~Hinderer.
\newblock A review of the {GGP} network and scientific challenges.
\newblock {\em Journal of Geodynamics}, 48(3-5):299--304, 2009.

\bibitem{Crossley2010}
D.~Crossley and J.~Hinderer.
\newblock {GGP} ({Global Geodynamics Project}): An international network of
  superconducting gravimeters to study time-variable gravity.
\newblock In {\em Gravity, Geoid and Earth Observation}, pages 627--635.
  Springer Berlin Heidelberg, 2010.

\bibitem{Horowitz2020a}
C.~J. Horowitz and R.~Widmer-Schnidrig.
\newblock Gravimeter search for compact dark matter objects moving in the
  earth.
\newblock {\em Physical Review Letters}, 124(5):051102, 2020.

\bibitem{Hu2020}
W.~Hu, M.~M. Lawson, D.~Budker, N.~L. Figueroa, D.~F.~Jackson Kimball, A.~P.
  Mills, and C.~Voigt.
\newblock A network of superconducting gravimeters as a detector of matter with
  feeble nongravitational coupling.
\newblock {\em The European Physical Journal D}, 74(6), 2020.

\bibitem{Mcnally2020}
R.~L. McNally and T.~Zelevinsky.
\newblock Constraining domain wall dark matter with a network of
  superconducting gravimeters and {LIGO}.
\newblock {\em The European Physical Journal D}, 74:1--6, 2020.

\bibitem{Liebling2012}
S.~L. Liebling and C.~Palenzuela.
\newblock Dynamical boson stars.
\newblock {\em Living Reviews in Relativity}, 15(1), 2012.

\bibitem{Giudice2016}
G.~F. Giudice, M.~McCullough, and A.~Urbano.
\newblock Hunting for dark particles with gravitational waves.
\newblock {\em Journal of Cosmology and Astroparticle Physics},
  2016(10):001--001, 2016.

\bibitem{Kimball2018}
D.{\hspace{0.167em}}F.~Jackson Kimball, D.~Budker, J.~Eby, M.~Pospelov,
  S.~Pustelny, T.~Scholtes, Y.{\hspace{0.167em}}V. Stadnik, A.~Weis, and
  A.~Wickenbrock.
\newblock Searching for axion stars and {Q}-balls with a terrestrial
  magnetometer network.
\newblock {\em Physical Review D}, 97(4), 2018.

\bibitem{Yang2017}
F.~Yang, M.~Su, and Y.~Zhao.
\newblock Dark matter annihilation from nearby ultra-compact micro halos to
  explain the tentative excess at 1.4 {TeV} in {DAMPE} data.
\newblock {\em arXiv preprint arXiv:1712.01724}, 2017.

\bibitem{Niikura2019}
H.~Niikura, M.~Takada, N.~Yasuda, R.~H. Lupton, T.~Sumi, S.~More, T.~Kurita,
  S.~Sugiyama, A.~More, M.~Oguri, and M.~Chiba.
\newblock Microlensing constraints on primordial black holes with subaru/{HSC}
  andromeda observations.
\newblock {\em Nature Astronomy}, 3(6):524--534, 2019.

\bibitem{Note1}
The feasibility of different capture scenarios is still up for debate. This is
  a non-trivial matter, especially considering that the CDOs would be
  virialized, so only a small fraction of them would be expected to have a
  velocity similar to that of the Earth (ie. within Earth's escape velocity).

\bibitem{Ostriker1999}
E.~C. Ostriker.
\newblock Dynamical friction in a gaseous medium.
\newblock {\em The Astrophysical Journal}, 513(1):252--258, 1999.

\bibitem{Note2}
A good way to understand this is to work in the Earth's rotating frame that
  rotates with the angular frequency of $2\pi /$day around the axis passing
  through the north and south poles. In this case, a CDO with a circular orbit
  in the equator plane will obtain a $\pm 2\pi /$day shift depending if it goes
  against (+) or with (-) the rotation of the Earth, and linear orbits along
  the south-north pole axis will remain unshifted. In general, a CDO's orbit
  can have components of these three cases, and therefore, the overall spectrum
  will generally have three components as well.

\bibitem{Note5}
Level 3 products are available for 26 stations and 36 sensors processed at 1
  min sampling. Level 3 data are the calibrated gravity residuals that have
  been corrected for: solid Earth tides, ocean tidal loading, atmospheric
  loading, polar motion and length‐of‐day induced gravity changes, and
  instrumental drift \cite{Boy2019}.

\bibitem{Boy2019}
J.~P. Boy.
\newblock Description of the level 2 and level 3 {IGETS} data produced by
  {EOST}; https://isdc.gfz-potsdam.de/igets-data-base/documentation/, 2019.

\bibitem{Ferreira2020}
E.~G.~M. Ferreira.
\newblock Ultra-light dark matter.
\newblock {\em arXiv preprint arXiv:2005.03254}, 2020.

\bibitem{Note4}
Gilad Perez and Abhishek Banerjee, personal communication.

\bibitem{Roberts2017}
B.~M. Roberts, G.~Blewitt, C.~Dailey, M.~Murphy, M.~Pospelov, A.~Rollings,
  J.~Sherman, W.~Williams, and A.~Derevianko.
\newblock Search for domain wall dark matter with atomic clocks on board global
  positioning system satellites.
\newblock {\em Nature communications}, 8(1):1--9, 2017.

\bibitem{Colella1975}
R.~Colella, A.~W. Overhauser, and S.~A. Werner.
\newblock Observation of gravitationally induced quantum interference.
\newblock {\em Physical Review Letters}, 34(23):1472, 1975.

\bibitem{Tino2020}
G.~M. Tino.
\newblock Testing gravity with cold atom interferometry: Results and prospects.
\newblock {\em Quantum Science and Technology}, 2020.

\bibitem{Elder2016}
B.~Elder, J.~Khoury, P.~Haslinger, M.~Jaffe, H.~M{\"u}ller, and P.~Hamilton.
\newblock Chameleon dark energy and atom interferometry.
\newblock {\em Physical Review D}, 94(4):044051, 2016.

\bibitem{Chiow2018}
S.~Chiow and N.~Yu.
\newblock Multiloop atom interferometer measurements of chameleon dark energy
  in microgravity.
\newblock {\em Physical Review D}, 97(4):044043, 2018.

\bibitem{Chiow2020}
S.~Chiow and N.~Yu.
\newblock Constraining symmetron dark energy using atom interferometry.
\newblock {\em Physical Review D}, 101(8):083501, 2020.

\bibitem{Kasevich1992}
M.~Kasevich and S.~Chu.
\newblock Measurement of the gravitational acceleration of an atom with a
  light-pulse atom interferometer.
\newblock {\em Applied Physics B}, 54(5):321--332, 1992.

\bibitem{Fang2016}
B.~Fang, I.~Dutta, P.~Gillot, D.~Savoie, J.~Lautier, B.~Cheng, C.~L.~G. Alzar,
  R.~Geiger, S.~Merlet, F.~P. Dos~Santos, and A.~Landragin.
\newblock Metrology with atom interferometry: Inertial sensors from laboratory
  to field applications.
\newblock {\em gravitational waves}, 8:9, 2016.

\bibitem{Freier2016}
C.~Freier, M.~Hauth, V.~Schkolnik, B.~Leykauf, M.~Schilling, H.~Wziontek,
  H.~Scherneck, J.~M{\"u}ller, and A.~Peters.
\newblock Mobile quantum gravity sensor with unprecedented stability.
\newblock In {\em Journal of Physics: Conference Series}, volume 723, page
  012050, 2016.

\bibitem{Karcher2018}
R.~Karcher, A.~Imanaliev, S.~Merlet, and F.~P. Dos~Santos.
\newblock Improving the accuracy of atom interferometers with ultracold
  sources.
\newblock {\em New Journal of Physics}, 20(11):113041, 2018.

\bibitem{Heine2020}
N.~Heine, J.~Matthias, M.~Sahelgozin, W.~Herr, S.~Abend, L.~Timmen,
  J.~M{\"u}ller, and E.~M. Rasel.
\newblock A transportable quantum gravimeter employing delta-kick collimated
  bose--einstein condensates.
\newblock {\em The European Physical Journal D}, 74(8):1--8, 2020.

\bibitem{Peters1999}
A.~Peters, K.~Y. Chung, and S.~Chu.
\newblock Measurement of gravitational acceleration by dropping atoms.
\newblock {\em Nature}, 400(6747):849--852, 1999.

\bibitem{Asenbaum2020}
P.~Asenbaum, C.~Overstreet, M.~Kim, J.~Curti, and M.~A. Kasevich.
\newblock Atom-interferometric test of the equivalence principle at the 10-12
  level, 2020.

\bibitem{Bonnin2013}
A.~Bonnin, N.~Zahzam, Y.~Bidel, and A.~Bresson.
\newblock Simultaneous dual-species matter-wave accelerometer.
\newblock {\em Phys. Rev. A}, 88:043615, 2013.

\bibitem{Zhou2015}
L.~Zhou, S.~Long, B.~Tang, X.~Chen, F.~Gao, W.~Peng, W.~Duan, J.~Zhong,
  Z.~Xiong, J.~Wang, Y.~Zhang, and M.~Zhan.
\newblock Test of equivalence principle at 10$^{-8}$ level by a dual-species
  double-diffraction raman atom interferometer.
\newblock {\em Physical Review Letters}, 115(1), 2015.

\bibitem{Schlippert2014}
D.~Schlippert, J.~Hartwig, H.~Albers, L.~L. Richardson, C.~Schubert, A.~Roura,
  W.~P. Schleich, W.~Ertmer, and E.~M. Rasel.
\newblock Quantum test of the universality of free fall.
\newblock {\em Phys. Rev. Lett.}, 112:203002, 2014.

\bibitem{Tarallo2014}
M.~G. Tarallo, T.~Mazzoni, N.~Poli, D.~V. Sutyrin, X.~Zhang, and G.~M. Tino.
\newblock Test of einstein equivalence principle for 0-spin and
  half-integer-spin atoms: Search for spin-gravity coupling effects.
\newblock {\em Phys. Rev. Lett.}, 113:023005, 2014.

\bibitem{Arvanitaki2018}
A.~Arvanitaki, P.~W. Graham, J.~M. Hogan, S.~Rajendran, and K.~Van~Tilburg.
\newblock Search for light scalar dark matter with atomic gravitational wave
  detectors.
\newblock {\em Phys. Rev. D}, 97:075020, 2018.

\bibitem{Graham2016}
P.~W. Graham, D.~E. Kaplan, J.~Mardon, S.~Rajendran, and W.~A. Terrano.
\newblock Dark matter direct detection with accelerometers.
\newblock {\em Phys. Rev. D}, 93:075029, 2016.

\bibitem{Geraci2016}
A.~A. Geraci and A.~Derevianko.
\newblock Sensitivity of atom interferometry to ultralight scalar field dark
  matter.
\newblock {\em Phys. Rev. Lett.}, 117:261301, 2016.

\bibitem{Hardman2016}
K.~S. Hardman, P.~J. Everitt, G.~D. McDonald, P.~Manju, P.~B. Wigley, M.~A.
  Sooriyabandara, C.~C.~N. Kuhn, J.~E. Debs, J.~D. Close, and N.~P. Robins.
\newblock Simultaneous precision gravimetry and magnetic gradiometry with a
  bose-einstein condensate: a high precision, quantum sensor.
\newblock {\em Physical review letters}, 117(13):138501, 2016.

\bibitem{Zhou2011}
L.~Zhou, Z.~Y. Xiong, W.~Yang, B.~Tang, W.~C. Peng, K.~Hao, R.~B. Li, M.~Liu,
  J.~Wang, and M.~S. Zhan.
\newblock Development of an atom gravimeter and status of the 10-meter atom
  interferometer for precision gravity measurement.
\newblock {\em General Relativity and Gravitation}, 43(7):1931--1942, 2011.

\bibitem{Overstreet2018}
C.~Overstreet, P.~Asenbaum, T.~Kovachy, R.~Notermans, J.~M. Hogan, and M.~A.
  Kasevich.
\newblock Effective inertial frame in an atom interferometric test of the
  equivalence principle.
\newblock {\em Physical review letters}, 120(18):183604, 2018.

\bibitem{Zhan2020}
M.~Zhan, J.~Wang, W.~Ni, D.~Gao, G.~Wang, L.~He, R.~Li, L.~Zhou, X.~Chen,
  J.~Zhong, B.~Tang, Z.~Yao, L.~Zhu, Z.~Xiong, S.~Lu, G.~Yu, Q.~Cheng, M.~Liu,
  Y.~Liang, P.~Xu, X.~He, M.~Ke, Z~Tan, and J.~Luo.
\newblock {ZAIGA: Z}haoshan long-baseline atom interferometer gravitation
  antenna.
\newblock {\em International Journal of Modern Physics D}, 29(04):1940005,
  2020.

\bibitem{Canuel2018}
B.~Canuel, A.~Bertoldi, L.~Amand, E.~P. Di~Borgo, T.~Chantrait, C.~Danquigny,
  M.~D. {\'A}lvarez, B.~Fang, A.~Freise, R.~Geiger, J.~Gillot, S.~Henry,
  J.~Hinderer, D.~Holleville, J.~Junca, G.~Lefèvre, M.~Merzougui, N.~Mielec,
  T.~Monfret, S.~Pelisson, M.~Prevedelli, S.~Reynaud, I.~Riou, Y.~Rogister,
  S.~Rosat, E.~Cormier, A.~Landragin, W.~Chaibi, S.~Gaffet, and P.~Bouyer.
\newblock Exploring gravity with the miga large scale atom interferometer.
\newblock {\em Scientific Reports}, 8(1):1--23, 2018.

\bibitem{Graham2017}
P.~W. Graham, J.~M. Hogan, M.~A. Kasevich, S.~Rajendran, and R.~W. Romani.
\newblock Mid-band gravitational wave detection with precision atomic sensors.
\newblock {\em arXiv preprint arXiv:1711.02225}, 2017.

\bibitem{Badurina2020}
L.~Badurina, E.~Bentine, D.~Blas, K.~Bongs, D.~Bortoletto, T.~Bowcock,
  K.~Bridges, W.~Bowden, O.~Buchmueller, C.~Burrage, J.~Coleman, G.~Elertas,
  J.~Ellis, C.~Foot, V.~Gibson, M.~G. Haehnelt, T.~Harte, S.~Hedges, R.~Hobson,
  M.~Holynski, T.~Jones, J.~March-Russell, C.~McCabe, D.~Newbold, B.~Sauer,
  U.~Schneider, I.~Shipsey, Y.~Singh, M.~A. Uchida, T.~Valenzuela, M.~van~der
  Grinten, V.~Vaskonen, J.~Vossebeld, D.~Weatherill, and I.~Wilmut.
\newblock {AION}: an atom interferometer observatory and network.
\newblock {\em Journal of Cosmology and Astroparticle Physics}, 2020(05):011,
  2020.

\bibitem{Bouyer2018}
P.~Bouyer.
\newblock {MIGA and ELGAR}: new perspectives for low frequency gravitational
  wave observation using atom interferometry. 2018.

\bibitem{Zhitnitsky2003}
A.~R. Zhitnitsky.
\newblock `{N}onbaryonic’ dark matter as baryonic colour superconductor.
\newblock {\em Journal of Cosmology and Astroparticle Physics},
  2003(10):010–010, 2003.

\bibitem{Flambaum2019}
V.~V. Flambaum and A.~R. Zhitnitsky.
\newblock Primordial lithium puzzle and the axion quark nugget dark matter
  model.
\newblock {\em Physical Review D}, 99(2):023517, 2019.

\bibitem{Raza2018}
N.~Raza, L.~Van~Waerbeke, and A.~Zhitnitsky.
\newblock Solar corona heating by axion quark nugget dark matter.
\newblock {\em Physical Review D}, 98(10):103527, 2018.

\bibitem{Vandevender2020}
J.~P. VanDevender, R.~G. Schmitt, N.~McGinley, A.~P. VanDevender, P.~Wilson,
  D.~Dixon, H.~Auer, and J.~McRae.
\newblock Results of search for magnetized quark-nugget dark matter from radial
  impacts on earth.
\newblock {\em arXiv preprint arXiv:2007.04826}, 2020.

\bibitem{Ge2020}
S.~Ge, H.~Rachmat, M.~S.~R. Siddiqui, L.~Van~Waerbeke, and A.~Zhitnitsky.
\newblock X-ray annual modulation observed by {XMM-N}ewton and axion quark
  nugget dark matter.
\newblock {\em arXiv preprint arXiv:2004.00632}, 2020.

\bibitem{Budker2020}
D.~Budker, V.~V. Flambaum, and A.~Zhitnitsky.
\newblock Axion quark nuggets. skyquakes and other mysterious explosions.
\newblock {\em arXiv preprint arXiv:2003.07363}, 2020.

\bibitem{Witten1984}
E.~Witten.
\newblock Cosmic separation of phases.
\newblock {\em Physical Review D}, 30(2):272, 1984.

\bibitem{Madsen1998}
J.~Madsen.
\newblock Physics and astrophysics of strange quark matter.
\newblock {\em Lecture Notes in Physics}, page 162–203, 1998.

\bibitem{Anderson2003}
D.~P. Anderson, E.~T. Herrin, V.~L. Teplitz, and I.~M. Tibuleac.
\newblock Unexplained sets of seismographic station reports and a set
  consistent with a quark nugget passage.
\newblock {\em Bulletin of the Seismological Society of America},
  93(6):2363--2374, 2003.

\bibitem{Herrin2006}
E.~T. Herrin, D.~C. Rosenbaum, and V.~L. Teplitz.
\newblock Seismic search for strange quark nuggets.
\newblock {\em Physical Review D}, 73(4):043511, 2006.

\bibitem{Green2012}
A.~M. Green.
\newblock Astrophysical uncertainties on direct detection experiments.
\newblock {\em Modern Physics Letters A}, 27(03):1230004, 2012.

\bibitem{Lawson2019}
K.~Lawson, X.~Liang, A.~Mead, M.~S.~R. Siddiqui, L.~Van~Waerbeke, and
  A.~Zhitnitsky.
\newblock Gravitationally trapped axions on the earth.
\newblock {\em Physical Review D}, 100(4):043531, 2019.

\bibitem{Budker2020a}
D.~Budker, V.~V. Flambaum, X.~Liang, and A.~Zhitnitsky.
\newblock Axion quark nuggets and how a global network can discover them.
\newblock {\em Physical Review D}, 101(4):043012, 2020.

\bibitem{Nakata2019}
N.~Nakata, L.~Gualtieri, and A.~Fichtner.
\newblock {\em Seismic ambient noise}.
\newblock Cambridge University Press, 2019.

\bibitem{Note6}
Simulations currently being worked on in collaboration with S. Yang, A.
  Zhitnitsky, V. Flambaum and C. Sens-Sch\"onfelder.

\bibitem{Note3}
Matter and antimatter could have been created in the same amount, so the total
  baryon charge of the Universe would be zero at all times, but more antimatter
  could be trapped inside nuggets (in contrast to the conventional baryogenesis
  paradigm where more matter than antimatter was produced in the early
  Universe). The excess of anti-nuggets to nuggets could arise from strong CP
  violation in the QCD epoch.

\bibitem{Kasevich1991}
M.~Kasevich and S.~Chu.
\newblock Atomic interferometry using stimulated raman transitions.
\newblock {\em Physical review letters}, 67(2):181, 1991.

\bibitem{Rosi2018}
G.~Rosi, A.~Vicer{\'e}, L.~Cacciapuoti, G.~D’Amico, L.~Hu, M.~Jain, N.~Poli,
  L.~Salvi, F.~Sorrentino, E.~Wang, et~al.
\newblock Detecting gravitational waves with atomic sensors.
\newblock {\em Il Nuovo Cimento}, 100(130):41, 2018.

\bibitem{Snadden1998}
M.~J. Snadden, J.~M. McGuirk, P.~Bouyer, K.~G. Haritos, and M.~A. Kasevich.
\newblock Measurement of the earth's gravity gradient with an atom
  interferometer-based gravity gradiometer.
\newblock {\em Phys. Rev. Lett.}, 81:971--974, Aug 1998.

\end{thebibliography}


\end{document}